\documentclass[letter]{IEEEtran}
\usepackage{amsmath}
\usepackage{graphicx}
\usepackage{caption}
\usepackage{epstopdf}
\usepackage{amsmath,amsfonts,amssymb,amsthm,epsfig,epstopdf,url,array}
\usepackage{graphicx}
\usepackage{color}
\usepackage{textcomp}
\newtheorem{theorem}{Theorem}

\usepackage{cite}
\usepackage{amsthm}
\usepackage{amsmath,amsthm}                                     
\usepackage{marvosym}                                 
\usepackage{array} 
\usepackage{algorithm}
\usepackage{algorithmic}
\usepackage{bm}

\usepackage{subfig}

\def\BibTeX{{\rm B\kern-.05em{\sc i\kern-.025em b}\kern-.08em
    T\kern-.1667em\lower.7ex\hbox{E}\kern-.125emX}}
\begin{document}
\title{Secure Semantic Communications: From Perspective of Physical Layer Security}
\author{
Yongkang Li,
Zheng Shi,
Han Hu,
Yaru Fu,
Hong Wang,
and Hongjiang Lei
\thanks{Yongkang Li and Zheng Shi are with the School of Intelligent Systems Science and Engineering, Jinan University, Zhuhai 519070, China (e-mails: A17200@stu2022.jnu.edu.cn, zhengshi@jnu.edu.cn).}
\thanks{Han Hu and Hong Wang are with Nanjing University of Posts and Telecommunications, Nanjing 210003, China (e-mails: {han\_h}@njupt.edu.cn, wanghong@njupt.edu.cn).}
\thanks{Yaru Fu is with the School of Science and Technology, Hong Kong Metropolitan University, Hong Kong SAR, China (e-mail: yfu@hkmu.edu.hk).}
\thanks{Hongjiang Lei is with Chongqing University of Posts and Telecommunications, Chongqing 400065, China (e-mail: leihj@cqupt.edu.cn).}
}
\markboth{Journal of \LaTeX\ Class Files,~Vol.~18, No.~9, September~2020}%
{How to Use the IEEEtran \LaTeX \ Templates}

\maketitle
\begin{abstract}
   Semantic communications have been envisioned as a potential technique that goes beyond Shannon paradigm. Unlike modern communications that provide bit-level security, the eavesdropping of semantic communications poses a significant risk of potentially exposing intention of legitimate user. To address this challenge, a novel deep neural network (DNN) enabled secure semantic communication (DeepSSC) system is developed by capitalizing on physical layer security. To balance the tradeoff between security and reliability, a two-phase training method for DNNs is devised. Particularly, Phase I aims at semantic recovery of legitimate user, while Phase II attempts to minimize the leakage of semantic information to eavesdroppers. The loss functions of DeepSSC in Phases I and II are respectively designed according to Shannon capacity and secure channel capacity, which are approximated with variational inference. Moreover, we define the metric of secure bilingual evaluation understudy (S-BLEU) to assess the security of semantic communications. Finally, simulation results demonstrate that DeepSSC achieves a significant boost to semantic security particularly in high signal-to-noise ratio regime, despite a minor degradation of reliability.
\end{abstract}
\section{Introduction}
\IEEEPARstart{T}{he} past decades have witnessed the explosive growth of capability of communications in transmitting symbols/bits, such as up to 20 Gigabits per second in 5G. The capacity of state-of-the-art communication systems is infinitely close to the Shannon limits, which preclude to further development of modern communication techniques. To break through such a bottleneck, semantic communications have received extensive attentions recently \cite{noma}. The essence of semantic communications is based on prior knowledge to extract the meaning of raw data by discarding the trivial information that has the least impact on semantic expression. Nowadays, the booming development of deep learning and natural language processing (NLP) has sparked of immense potential of semantic communications to accommodate a more advanced paradigm of communication \cite{xie2021deep,bourtsoulatze2019deep,9953110}. 
 To name a few, a deep learning enabled end-to-end communication system was proposed in \cite{xie2021deep}, where the architecture of Transformer was adopted to extract textual semantics.
Regarding image transmission, the convolutional neural network was used to preserve meaning of image \cite{bourtsoulatze2019deep}. Furthermore, the contextual nonlinear Transform was invoked to extract semantic features across temporally correlated frames for video transmission \cite{9953110}. 

Most of previous literature focus on reliability and effectiveness of semantic communications. More specifically, in \cite{jiang2022deep}, hybrid automatic repeat request (HARQ) was exploited to reduce the semantic errors. To better adapt to dynamic channel conditions, an improved semantic encoding solution with multibit length selection was designed for HARQ with incremental knowledge in \cite{zhou2022adaptive}. By introducing transformer-based knowledge extractor, the semantic decoding capability was enhanced in \cite{wang2023knowledge}. Besides, Hu {\em et al.} in \cite{hu2022one} concerned with multi-user semantic communications, where distinct semantic features were utilized to distinguish different users at the receiver. 
Aside from the aforementioned two fundamental metrics, i.e., reliability and effectiveness, the security becomes of ever-increasing importance and is conceived as the key warranty to the success of 6G \cite{nguyen2021security}. 
Due to the broadcast nature of radio propagation, wireless communications are susceptible to eavesdropping. However, investigations into wireless secure semantic communications are still in their infancy. Furthermore, it is noteworthy that the leakage of semantic information might lead to the exposure of intention to unintended recipients, consequently yields more serious security threats compared to bit/symbol-level based communication systems.
To overcome this issue, only a few existing efforts have taken possible safeguards to offer the security of semantic communications. For example, the semantic information is encrypted with a secret key for security preserving \cite{luo2023encrypted}. In addition, a joint-source-channel autoencoder was proposed in \cite{zhang2023wireless} for efficient semantic meaning extraction of image, where secure mean squared-error was used as loss function to govern efficiency-privacy trade-off. {{Most of existing works considered the security of semantic communications from upper layer of network, while ignored the risk caused by key leakage \cite{luo2023encrypted} and the effect of fading channels \cite{zhang2023wireless}. 
This motivates us to apply physical layer security (PLS) to realize secure semantic communications. PLS essentially reaps the benefits of the intrinsic randomness and dynamics of wireless channels to protect data confidentiality. Indeed, there is a consensus on the pivotal role of PLS in the provision of robust security for 6G \cite{nguyen2021security}.}} Hence, we focus on PLS-empowered secure semantic communications. 

The main contributions can be summarized as follows.
\begin{itemize}
    \item A deep neural network (DNN) enabled secure semantic communication (DeepSSC) system is developed. The training process of the DeepSSC is separated into two phases to trade off security with reliability. In particular, Phases I and II are designed to provide reliable and secure semantic communications, respectively. 
    \item {Secure semantic communications are enabled by virtue of PLS.} Towards this end, the loss functions in two phases are respectively designed according to Shannon capacity and secure channel capacity, which are approximated with variational inference for tractability.
    \item {The metric of secure bilingual evaluation understudy (S-BLEU) is proposed to assess the security of semantic communications.} Simulation results verify that the DeepSSC tremendously improves semantic security especially at high signal-to-noise ratio (SNR), albeit at the expense of a slight reduction in reliability.
\end{itemize}

The rest of this letter is outlined as follows. In Section II, we introduce the system model and performance metrics. The DeepSSC is proposed in Section III. Section IV presents numerical results. Section V finally concludes this letter.

\section{System Model and Performance Metrics}
This section introduces a secure semantic communication system, followed by the metrics of reliability and security.
\subsection{Secure Semantic Communication System}
\subsubsection{Wire-Tap Channel Model}
\begin{figure}[htbp]
    \centering
    \includegraphics[width=3in,height=0.75in]{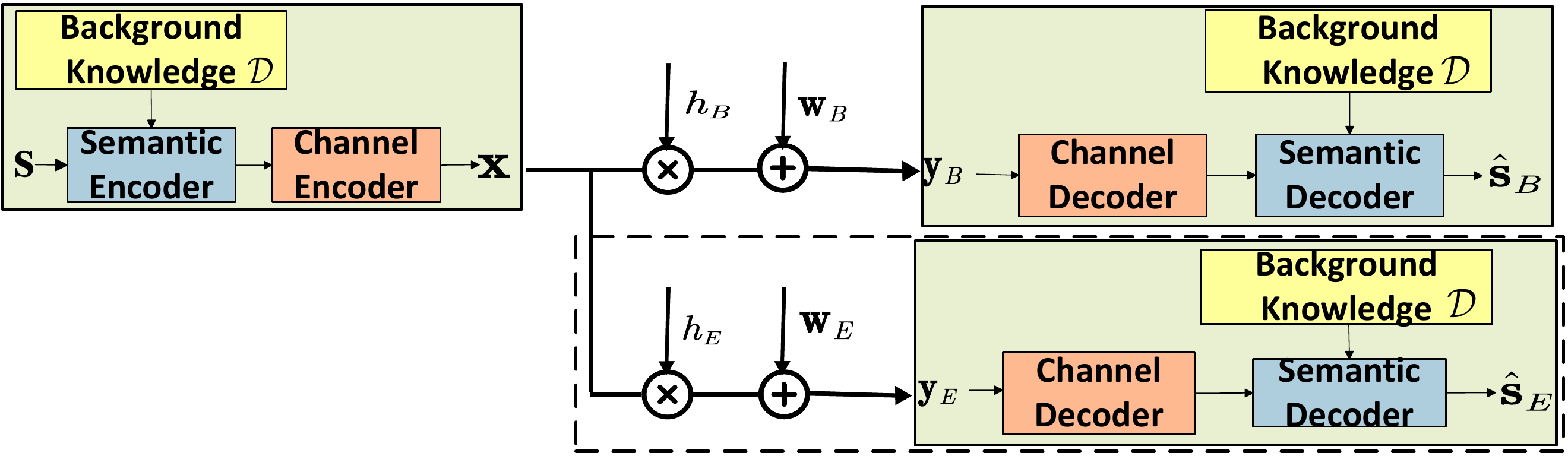}
    \caption{The framework of DeepSSC.}
    \label{system}
\end{figure}

Fig. \ref{system} depicts a classical wire-tap channel model, which consists of a transmitter (i.e., Alice), a legitimate receiver (i.e., Bob), and an eavesdropper (i.e., Eve). In particular, Alice sends confidential semantic information to Bob through main channel. Meanwhile, Eve overhears the emitted semantic signal. Both Alice and Bob extract and reconstruct semantic information on the basis of their background knowledge. 
Without the background knowledge of Alice, Eve obviously has difficulty understanding the semantics from the eavesdropped signal. In this letter, we only consider the PLS by disregarding the effect of different background knowledge sets. Accordingly, we assume that the background knowledge $\mathcal D$ of Alice is assumed to be publicly known to both Bob and Eve.
\vspace{-10pt}
\subsubsection{System Description}
{As a representative technique of implicit reasoning, deep learning is able to offer effective semantic extraction and end-to-end communications.} As illustrated in Fig. \ref{system}, the secure semantic communication system is realized by using deep neural network (DNN)-based framework, named as DeepSSC. We denote the source sentence as ${\bf{s}}=[w_1,w_2,...w_L]$, where $w_l$ represents the $l$-th word in ${\bf{s}}$. Prior to the delivery of $\bf s$, Alice first encodes $\bf s$ through a semantic encoder that extracts its involved semantic information. The generated semantic sequence is then sent to the channel encoder, which introduces redundant information to ensure the reliable transmission of the semantic information over noisy channels. More precisely, the encoded symbol sequence ${\bf{x}}\in \mathbb C^{M\times1}$ is mathematically expressed as
\begin{equation}
   {\bf{x}} = {T^C}({T^S}(\bf{s};\bf{\bm{\alpha}} );\bf{\bm{\beta}} ),
\end{equation}
where $M$ is the length of the symbol sequence, ${T^S}(\cdot;\bf{\bm{\alpha}} )$ and ${T^C}(\cdot;\bf{\bm{\beta}} )$ denote the DNN-enabled semantic encoder and the DNN-enabled channel encoder, respectively, $\bf{\bm{\alpha}}$ and $\bf\bm{\beta} $ correspond to the parameter sets of DNNs. 

Denote by ${\bf{y}}_B, {\bf{y}}_E \in \mathbb C^{M\times1}$ the corrupted signals received at Bob and Eve, respectively. By assuming block fading channels, the received signal is written as
\begin{align}\label{model}
{\bf y}_{\kappa} = \sqrt{P}{{h_{\kappa}}}{\bf x} + {{\bf w}_{\kappa}},\,\kappa\in \{B,E\},
\end{align}
where $P$ represents the transmit power, ${\bf{w_{\kappa}}}$ stands for zero-mean additive white Gaussian noise (AWGN) with variance $\mathcal N$, ${h_{\kappa}}$ refers to the channel coefficient. 
After receiving the signal ${\bf y}_{\kappa}$ at Bob and Eve, the receiver first reconstructs the transmitted semantic symbols via the channel decoder. The semantic decoder is then in charge of converting the semantic information into the original sentence. Therefore, the recovered sentence ${{\bf{s_\kappa}}}$ can be represented as
\begin{align}
{\hat{\bf{s}}_\kappa} = R_\kappa^S(R_\kappa^C({\bf{y_\kappa}};{\bf{\bm{\chi}} _\kappa});{\bf{\bm{\delta}} _\kappa}),\,\kappa\in \{B,E\},
\end{align}
where $R_\kappa^C(\cdot;{\bf{\bm{\chi}} _\kappa})$ denotes the channel decoder with the parameter set ${\bf{\bm{\chi}} _\kappa}$ and $R_\kappa^S(\cdot;{\bf{\bm{\delta}} _\kappa})$ refers to the semantic decoder with the parameter set ${\bf{\bm{\delta}} _\kappa}$.
\subsection{Performance Metrics}
{Unlike conventional bit-wise communications, performance metrics such as bit error rate, outage probability, etc., are no longer applicable to semantic communications, because semantic communications concentrate on the successful delivery of the meanings of data rather than the bit sequence.} In order to evaluate the reliability of semantic communications, BLEU score is used. In addition, S-BLEU score is proposed to investigate the security of semantic communications. These two metrics are introduced as follows.
\subsubsection{BLEU Score}
{In natural language processing (NLP), the BLEU score is a frequently adopted metric to measure the word-level similarity between two sentences \cite{xie2021deep}}.  Herein, the BLEU score is employed to evaluate the reliability of semantic communications. The BLEU score between $\bf s$ and $\hat{\bf s}_\kappa$ is defined as
\begin{equation}\label{eqn:bleu_def}
    \log {\rm {BLEU}} = 
 [1-{{{l_\mathbf{s}}}}/{l_{\hat{\bf{s}}_\kappa}}]^- + \sum\nolimits_{n = 1}^{\mathfrak N} {{u_n}\log {f_n}}, 
\end{equation}
where $[x]^-=\min \left\{x,0\right\}$, $l_\mathbf{s}$ and $l_{{\hat{\bf{s}}_\kappa}}$ represent the lengths of the original sentence $\mathbf{s}$ and the recovered sentence ${\hat{\bf{s}}_\kappa}$, respectively, $u_n$ is the weights of $n$-grams, and $f_n$ indicates the $n$-grams score that can be calculated as
\begin{equation}\label{eqn:bleu_e}
    f_{n}=\frac{\sum_{k} \min \left(C_{k}({\bf{\hat{s}}_{\kappa}}), C_{k}(\mathbf{s})\right)}{\sum_{k}  C_{k}({\bf{\hat{s}}_{\kappa}})},
\end{equation}
where $C_{k}(\cdot)$ is the function counting the frequency of the $k$-th element in $n$-grams and $k \in [1,L-n+1]$.

\subsubsection{S-BLEU Score}
Similarly to BLEU score, secure BLEU (S-BLEU) score is defined to measure the security degree that how many words have not been eavesdropped. Accordingly, S-BLEU score can be expressed as 
\begin{equation}\label{eqn:s-bleu-def}
    \log {\text {S-BLEU}} = 
[1 - {{{l_{\bf s}}}}/{l_{\hat{\mathbf{s}}_B}}]^- + \sum\nolimits_{n = 1}^N {{u_n}\log {\bar f_n}} 
\end{equation}
where $[1 - {l_{{\hat{\mathbf{s}}_B}}}/{{{l_{\bf s}}}}]^-$ is a short sentence penalty term 
and $\bar f_n$ stands for the $n$-grams secure score. In analogous to \eqref{eqn:bleu_e}, $\bar f_n$ can be computed as
\begin{equation}
    \bar f_{n}=\frac{\sum_{k} \min \left(  [C_{k}({\hat{\mathbf{s}}_B})-C_{k}({\hat{\mathbf{s}}_E})]^+,   C_{k}(\mathbf{s})\right)}{\sum_{k} C_{k}({\hat{\mathbf{s}}_B})},
\end{equation}
where the notation $[x]^+=\max\{x,0\}$. For example, if ``weather is good today'' is sent by Alice, Bob receives ``weather is nice today'' and Eve eavesdrops ``weather good''. According to \eqref{eqn:bleu_def} and \eqref{eqn:s-bleu-def}, the BLEU and S-BLEU scores are ${3/4}$ and ${1/2}$, respectively.
\vspace{-10pt}

\section{Secure Semantic Communications}
This section first formulates the problem of secure semantic communications, with which a deep learning enabled model, i.e., DeepSSC, is designed. At last, the training algorithm is proposed to update the neural networks. 
\subsection{Problem Formulation}
In this letter, our goal is to minimize semantic errors at Bob while preventing Eve from intercepting confidential semantic information as much as possible. 
Moreover, in order to manage the tradeoff between the reliability and the security, we develop a two-phase training approach for a joint design of source and channel codings, in which the first and the second phases aim at the reliability assurance and the security guarantee, respectively.
\subsubsection{Phase I of Reliability Assurance}
In particular, the first phase is devoted to preserving as much semantic information. In order to retrieve the sentence $s$ from the observation $\hat {\bf y}_\kappa$, the framework of joint source-channel coding (JSCC) can be established to maximize the mutual information of the source sentence, $\bf s$, and the received symbols, ${\bf y}_\kappa$, i.e.,\footnote{{In this letter, Eve aims to achieve the minimum semantic errors through maximizing the channel capacity of eavesdropping channel. This actually corresponds to the worst case according to Phase II training.}}
\begin{equation}
\begin{array}{*{20}{c}}\label{Ixy}
{\boldsymbol{\phi}_\kappa } = {\mathop {\arg \max }\limits_{\bm{\alpha} ,\bm{\beta} ,\bm{\chi}_\kappa ,\bm{\delta}_\kappa}  }&{\mathbb E_{{p(h_\kappa)}}\left\{{I({\bf{s}};\left. {\bf y}_\kappa\right|h_\kappa)}\right\}},
\end{array}
\end{equation}
where $\boldsymbol{\phi}_\kappa \triangleq \{\bm{\alpha} ,\bm{\beta} ,\bm{\chi}_\kappa ,\bm{\delta}_\kappa\}$, $\kappa\in \{B,E\}$, and $I({\bf x};{\bf y}) =\mathbb E_{\bf x,y}\{\log p(x|y)/p(x)\}$. 
{The JSCC is able to offer real end-to-end semantic communications and exploit the high-level semantic features, albeit with the drawbacks of high training overhead, poor robustness, and poor generalization ability.} However, it is not easy to directly optimize the objective function \eqref{Ixy} in practice. In addition, it does not take into account the decoders. For tractability, we resort to an alternative approximation of \eqref{Ixy} based on variational inference, as given by the following theorem.
\begin{theorem}\label{theorem1}
A variational inference lower bound of $I({\bf{s}};\left. {\bf y}_\kappa\right|h_\kappa)$ is obtained as
\begin{align}\label{app}
&I({\bf{s}};\left. {{{\bf{y}}_\kappa }} \right|{h_\kappa }) =  - H(\left. {\bf{s}} \right|{{\bf{y}}_\kappa },{h_\kappa }) + H({\bf{s}})\notag\\
&=  {\mathbb E_{p\left( {\left. {{\bf{s}},{{\bf{y}}_\kappa }} \right|{h_\kappa }} \right)}}\left\{ {\log \left( {p(\left. {\bf{s}} \right|{{\bf{y}}_\kappa },{h_\kappa })\frac{{p(\left. {{\bf{\hat s}}_\kappa} \right|{{\bf{y}}_\kappa },{h_\kappa },{\boldsymbol{\phi} _\kappa })}}{{p(\left. {{\bf{\hat s}}_\kappa} \right|{{\bf{y}}_\kappa },{h_\kappa },{\boldsymbol{\phi} _\kappa })}}} \right)} \right\}\notag\\
&\quad+ H({\bf{s}})  \notag\\
& = {\mathbb E_{p\left( {\left. {{\bf{s}},{{\bf{y}}_\kappa }} \right|{h_\kappa }} \right)}}\left\{ {\log \frac{{p(\left. {\bf{s}} \right|{{\bf{y}}_\kappa },{h_\kappa })}}{{p(\left. {{\bf{\hat s}}_\kappa} \right|{{\bf{y}}_\kappa },{h_\kappa },{\boldsymbol{\phi} _\kappa })}}} \right\}\notag\\
& \quad + {\mathbb E_{p\left( {\left. {{\bf{s}},{{\bf{y}}_\kappa }} \right|{h_\kappa }} \right)}}\left\{ {\log p(\left. {{\bf{\hat s}}_\kappa} \right|{{\bf{y}}_\kappa },{h_\kappa },{\boldsymbol{\phi} _\kappa })} \right\}+ H({\bf{s}})  \notag\\
&\mathop  \ge \limits^{\left(a \right)}  H({\bf{s}}) + {\mathbb E_{p\left( {\left. {{\bf{s}},{{\bf{y}}_\kappa }} \right|{h_\kappa }} \right)}}\left\{ {\log p(\left. {{\bf{\hat s}}_\kappa} \right|{{\bf{y}}_\kappa },{h_\kappa },{\boldsymbol{\phi} _\kappa })} \right\} \notag\\
&\mathop  = \limits^{\left( b \right)}   H({\bf{s}})- {\mathbb E_{p\left( {\left. {{{\bf{y}}_\kappa }} \right|{h_\kappa }} \right)}}\left\{ {{{\cal L}_{{\rm{CE}}}}({\bf{s}},{\bf{\hat s}}_\kappa)} \right\}, 
\end{align}
where $H(\cdot)$ refers to the entropy functional, Step (a) holds by using the non-negative property of KL divergence, Step $(b)$ holds by virtue of the definition of cross entropy, and the cross entropy $ {{\cal L}_{{\rm{CE}}}}({\bf{s}},{\bf{\hat s}}_\kappa)$ of the semantic decoder ${p(\left. {\hat {\bf{s}}}_\kappa \right|{\bf y},h_\kappa, \boldsymbol{\phi}_\kappa )}$ relative to the true posterior ${p(\left. { {\bf{s}}}_\kappa \right|{\bf y},\boldsymbol{\phi}_\kappa )}$ is defined as
${{\cal L}_{{\rm{CE}}}}({\bf{s}},{\bf{\hat s}}_\kappa) =  - {\mathbb E_{p(\left. {\bf{s}} \right|{{\bf{y}}_\kappa },{h_\kappa })}}\left\{ {\log p(\left. {{\bf{\hat s}}_\kappa} \right|{{\bf{y}}_\kappa },{h_\kappa },{\boldsymbol{\phi}_\kappa })} \right\}.
$
\end{theorem}

According to Theorem \ref{theorem1}, the cross entropy ${{\cal L}_{{\rm{CE}}}}({\bf{s}}_\kappa,{\bf{\hat s}}_\kappa)$ is used as the loss function in the first phase. 
Similarly to \cite{xie2020lite}, back propagation is applied to update the parameters of neural networks. By taking the semantic encoder ${T^S}(\bf{s};\bf{\bm{\alpha}} )$ as an example, the gradient descent method is leveraged to update the parameter vector $\bm{\alpha}$ at the $t$-th iteration as $\bm{\alpha}(t)=\bm{\alpha}(t-1)-\eta \frac{{\partial {{\cal L}_{{\rm{CE}}}}}}{{\partial {\bm{\alpha} ^{\rm{T}}}}}$, where $\eta$ is the learning rate and the gradient $ \frac{{\partial {{\cal L}_{{\rm{CE}}}}}}{{\partial {\bm{\alpha} ^{\rm{T}}}}}$ is given by
\begin{align}
\frac{{\partial {{\cal L}_{{\rm{CE}}}}}}{{\partial {\bm{\alpha} ^{\rm{T}}}}} &= \frac{{\partial {{\cal L}_{{\rm{CE}}}}}}{{\partial {{{\bf{\hat s}}}_\kappa }^{\rm{T}}}}\frac{{\partial {{{\bf{\hat s}}}_\kappa }}}{{\partial {{\bf{y}}_\kappa }^{\rm{T}}}}\frac{{\partial {{\bf{y}}_\kappa }}}{{\partial {{\bf{x}}^{\rm{T}}}}}\frac{{\partial {\bf{x}}}}{{\partial {\bm{\alpha} ^{\rm{T}}}}} = h_{\kappa}\frac{{\partial {{\cal L}_{{\rm{CE}}}}}}{{\partial {{{\bf{\hat s}}}_\kappa }^{\rm{T}}}}\frac{{\partial {{{\bf{\hat s}}}_\kappa }}}{{\partial {{\bf{y}}_\kappa }^{\rm{T}}}}\frac{{\partial {\bf{x}}}}{{\partial {\bm{\alpha} ^{\rm{T}}}}}.
\end{align}

\subsubsection{Phase II of Security Guarantee}
In order to ensure secure semantic communications, the goal in the second phase is to minimize the leakage of semantic information to the eavesdropper from the perspective of physical layer. 
According to the fundamental principle of PLS, the ergodic secrecy capacity is written as {\cite{5352243}}
\begin{align}
  C_s = \mathbb E_{p(h_B,h_E)}\{[C_B-C_E]^+\},  
\end{align}
where 
$C_B$ and $C_E$ are the main channel capacity and the eavesdropping channel capacity, respectively. From the information-theoretical perspective, it follows that $C_B = I({\bf{s}};\left.{\bf y}_B\right|h_B)$ and $C_E = I( {\bf{s}};\left.{\bf y}_E\right|h_E)$.  
On this basis, $ C_s = \mathbb E _{p(h_B,h_E)}[{{ I( {\bf{s}};\left.{\bf y}_B\right|h_B)-I( {\bf{s}};\left.{\bf y}_E\right|h_E)}}]^+$ can be approximated with Theorem \ref{theorem1} as 
\begin{equation}\label{tal}
    {C}_{s} \approx -{\mathbb E_{p({\bf y}_B ,h_B,{\bf y}_E ,h_E)}}\left\{ [L_{\rm SSC}({\bf{s}},\hat {\bf{s}}_\kappa]^-)  \right\}. 
\end{equation}
where $L_{\rm SSC}({\bf{s}},\hat {\bf{s}}_\kappa) = \mathcal{L}_{\mathrm{CE}}( {\bf{s}},{ \hat{\bf{s}}_B})  - \mathcal{L}_{\mathrm{CE}}( {\bf{s}},{ \hat{\bf{s}}_E})$.
Accordingly, we adopt $L_{\rm SSC}({\bf{s}},\hat {\bf{s}}_\kappa)$ as the loss function in the second phase.

{It is noteworthy that the above two phases can be merged together as a whole. More specifically, it can be trained with a unified loss function, which is expressed as a weighted sum of ${{\cal L}_{{\rm{CE}}}}({\bf{s}},{\bf{\hat s}}_\kappa) $ and $L_{\rm SSC}({\bf{s}},\hat {\bf{s}}_\kappa)$ such that
\begin{equation}\label{eqn:inte}
    L_{\rm INT}({\bf{s}},\hat {\bf{s}}_\kappa) = (w_1+w_2)\mathcal{L}_{\mathrm{CE}}( {\bf{s}},{ \hat{\bf{s}}_B})  - w_2\mathcal{L}_{\mathrm{CE}}( {\bf{s}},{ \hat{\bf{s}}_E}),
    \end{equation}
    where $w_1$ and $w_2$ are the weights.}

\subsection{Model Design}
\begin{figure}[htb!]
    \centering
    \includegraphics[width=3in]{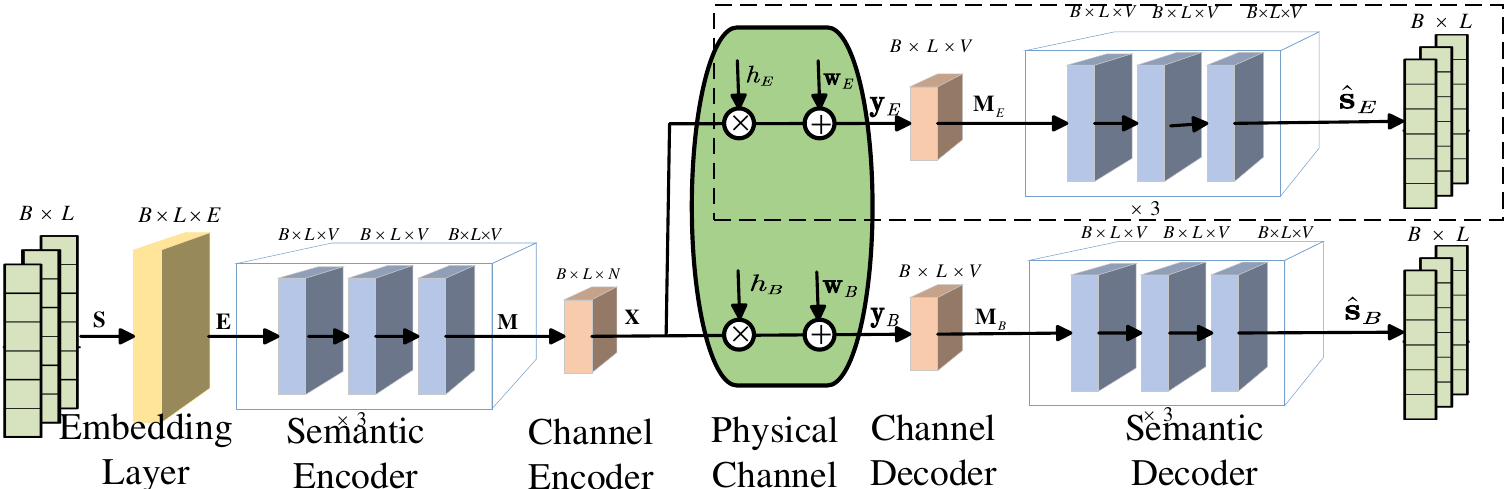}
    \caption{The schematic diagram of DeepSSC.}
    \label{semantic}
\end{figure}

{It should be noticed that the proposed DeepSSC does not have any specific requirements on the network architecture. Without loss of generality, the model in \cite{xie2021deep} is used in this letter.} The proposed deep learning model for DeepSSC is shown in Fig. \ref{semantic}. 
In particular, a small batch of sentences ${\bf{S}} \in \mathbb{R}^{B\times L}$ randomly drawn from a knowledge set $\mathcal{D}$ are used as input, where $B$ is the batch size and $L$ is the length of sentence. The embedding layer is subsequently responsible for representing each sentence with a dense word vector $\mathbf{E} \in \mathbb{R}^{B\times L\times E}$, where $E$ is the dimension of the word vector. The semantic encoder layer uses $\mathbf E$ as the input to obtain semantic information matrix $\mathbf{M}\in \mathbb{R}^{B\times L\times V}$, where $V$ is the output dimension of Transformer encoder layer. The semantic encoder commonly contains multiple Transformer encoding layers with each consisting of one self-attention sublayer and one feed-forward sublayer. As the key component of the Transformer, self-attention layer empowers the semantic encoder to extract the meaning of sentences, while feed-forward layer 
enhances the network fitting capability. To combat the signal distortion, the semantic information $\mathbf{M}$ is converted into transmitted symbols $\mathbf{X}\in \mathbb{R}^{B\times L\times N}$ by the channel encoder that is composed of multiple dense layers, where $N$ is the length of the symbols for each sentence.

The received symbols at Bob and Eve are denoted by $\mathbf{Y}$ and $\mathbf{Z}$, respectively. Since both Bob and Eve employ the same decoding model, the receiver structure at Bob is taken as an example to conserve space. With the corrupted symbols $\mathbf{Y}$, Bob recovers the semantic information matix $\mathbf{M }_B\in \mathbb{R}^{B\times L\times V}$ through a channel decoder layer, which consists of multiple dense layers. Subsequently, the source sentence is estimated as $\bf{S _B}$ through a transformer-enabled semantic decoder. The semantic decoder contains multiple Transformer decoding layers with each having three sublayers, namely, the self-attention sublayer, encoder-decoder attention sublayer, and feed-forward sublayer. 





\subsection{Training Algorithm}
The training process of the proposed DeepSSC is split into two phases. In particular, the DNNs at Alice, Bob, and Eve are trained in Phase I to achieve the minimum semantic errors. 
In Phase II, we continue to train the DNNs at Alice and Bob to cope with the eavesdropping of Eve.



The pseudocode of training phase I is displayed in Algorithm \ref{alg2}. At first, a small batch of sentences $\bf{S}$ are drawn from the knowledge set $\mathcal{D}$ and then encoded into semantic information matrix $\bf{M}$ through the semantic encoder ${T^S}(\bf{S};{\bf\bm{\alpha}} )$. Subsequently, passing $\mathbf{M}$ to the channel encoder ${T^C}(\bf{M};{\bf\bm{\beta}} )$  yields $\mathbf{X}$ that is transmitted over the fading channels. Bob receives the corrupted signal $\mathbf{Y}$, with which the semantic information matrix ${\bf M}_B$ is reconstructed via the channel decoder ${R^C_B}({\mathbf Y};{\bm{\chi}_B} )$. The source sentence $\mathbf{S}_B$ is therefore estimated with the semantic decoder ${R^S_B}({\mathbf M_B;\bm{\delta}_B )} $. As a consequence, stochastic gradient descent (SGD) is applied to update the parameters of DNNs between Alice and Bob by using ${{\cal L}_{{\rm{CE}}}}({\bf{s}}_\kappa,{\bf{\hat s}}_\kappa)$ as the loss function. Since Eve has the same structure of DNNs as Bob, transfer learning can be leveraged to update the parameters $\mathbf{\bm{\chi}}_E$ and $\mathbf{ \bm{\delta}}_E$ after loading the pre-trained DNNs at Bob. In Phase II, we freeze the channel and the semantic decoders (i.e., ${R^C_E}({\bf\cdot};{\bf\bm{\chi}_E} )$ and $ {R^S_E}({\bf\cdot};{\bf\bm{\delta}_E} )$) at Eve,  and then retrain ${T^S}({\bf\cdot};{\bf\bm{\alpha}} )$, ${T^C}({\bf\cdot};{\bf\bm{\beta}} )$, ${R^C_\kappa}({\bf\cdot};{\bf\bm{\chi}}_B )$, and ${R^S_B}({\bf\cdot};{\bf\bm{\delta}}_\kappa )$ by repeating Steps 1-11 in Algorithm \ref{alg2}. Whereas, $ L_{\rm SSC}({\bf{s}},\hat {\bf{s}}_\kappa)$ is chosen as the loss function in Phase II.


\begin{algorithm}
    \caption{Training Phase I}
    \label{alg2}
    \begin{algorithmic}[1]
    \STATE \textbf{Alice:} 
          \STATE \hspace{2em}Generate a small batch $\bf{S}$ from $\mathcal{D}$.
          \STATE \hspace{2em}${T^C}({T^S}(\bf{S};{\bf\bm{\alpha}} );{\bf\bm{\beta}} ) \to$ transmitted symbols $ \bf{X}$.
          \STATE \hspace{2em}Transmit $\bf{X}$ over fading channels.
    \STATE \textbf{Bob:}
          \STATE \hspace{2em}Receive $\bf{Y}$.
          \STATE \hspace{2em}${R^S_B}({{R^C_B}({\bf Y};{\bf\bm{\chi}_B} );\bm{\delta}_B )} \to$ estimated sentence ${\bf S_B} $.
          \STATE \hspace{2em}Compute loss function ${\mathbb E_{p\left( {\left. {{{\bf{y}}_\kappa }} \right|{h_\kappa }} \right)}}\left\{ {{{\cal L}_{{\rm{CE}}}}({\bf{s}},{\bf{\hat s}}_\kappa)} \right\}$.
          \STATE \hspace{2em}Update ${\bf\bm{\alpha}}$, ${\bf\bm{\beta}}$, ${\bf\bm{\chi}_B}$, $\bf \bm{\delta}_B$  with SGD. 
    \STATE \textbf{Eve:}
          \STATE \hspace{2em}Freeze ${T^S}({\bf\cdot};{\bf{\bm{\alpha}}} )$ and ${T^C}({\bf\cdot};{\bf\bm{\beta}} )$.
          \STATE \hspace{2em}Repeat Steps 6-9 to train the DNNs at Eve.

    \RETURN 
    ${T^S}({\bf\cdot};{\bf\bm{\alpha}} ), {T^C}({\bf\cdot};{\bf\bm{\beta}} ), 
    {R^C_\kappa}({\bf\cdot};{\bf\bm{\chi}}_B ), {R^S_B}({\bf\cdot};{\bf\bm{\delta}}_\kappa )$.
   
    \end{algorithmic}
     
\end{algorithm}
 \vspace{-20pt}

\begin{figure*}[t]
	\begin{minipage}[b]{.31\linewidth}
		\centering
		\includegraphics[width=\linewidth,height=3cm]{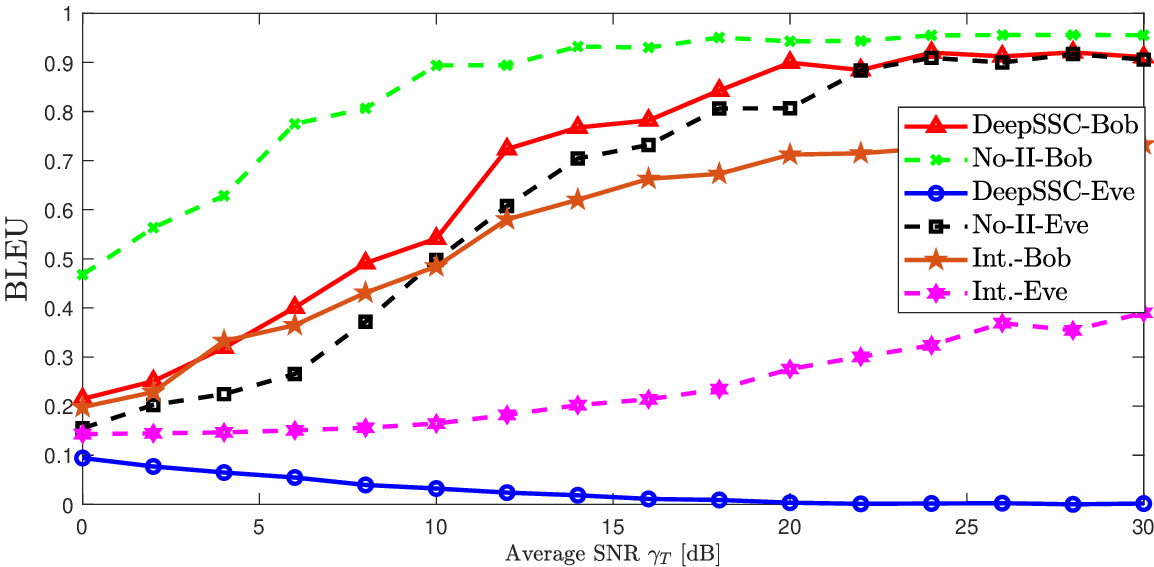}
		\caption{BLEU score of 1-gram versus  SNR $\gamma_T$.}
		\label{fig1}
	\end{minipage}
        \hfill
	\begin{minipage}[b]{.31\linewidth}
		\centering
		\includegraphics[width=\linewidth,height=3cm]{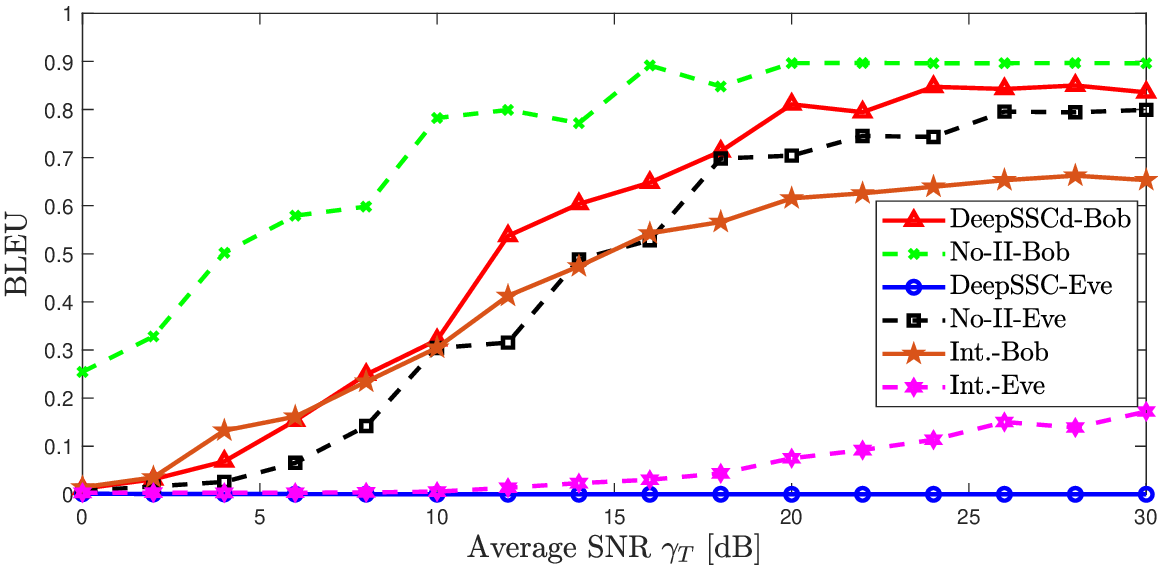}
		\caption{BLEU score of 3-grams versus  SNR $\gamma_T$.}
		\label{3_gram}
	\end{minipage}
        \hfill
	\begin{minipage}[b]{.31\linewidth}
		\centering
		\includegraphics[width=\linewidth,height=3cm]{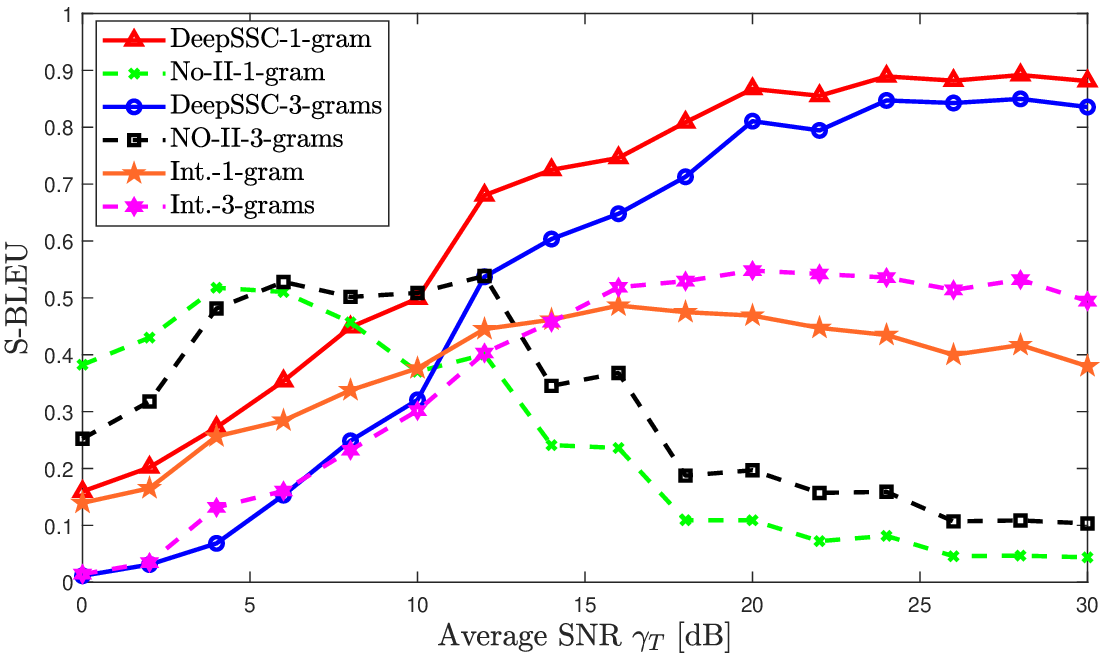}
		\caption{S-BLEU scores of 1-gram and 3-grams versus SNR $\gamma_T$.}
		\label{fig2}
	\end{minipage}
        \hfill
        \vspace{-15pt}
\end{figure*}
 
\section{Simulation Results}
This section presents experimental results for performance assessment of the proposed DeepSSC. {{In simulations, we assume that the carrier frequency is $f_c=1$ GHz and sub-6 GHz channels are frequently modeled as $h_\kappa = \mu {d_\kappa}^{-2} X_\kappa$, where $\mu=(c/(4\pi f_c))^2$, $d_\kappa$, and $X_\kappa$ denote the path loss at a reference distance, the communication distance, and the small-scale Rayleigh fading with ${\cal CN}(0,1)$, respectively \cite{2018pls}. The bandwidth is $B=20$ MHz, and noise power $\mathcal N=-174+10 \log_{10}(B)+N_f$ with noise figure $N_f=10$dB}. The distances from Alice to Bob and Eve are assumed to be $1$ km and $3$ km, respectively.} 
We employ the dataset of proceedings of European Parliament for training and testing \cite{koehn2005europarl}. The dataset comprises approximately 2 million sentences each with 4 to 30 words. The dataset is split into training data and testing data, wherein the testing data contains around 60,000 sentences and the rest of sentences are used for the training. Moreover, semantic encoder and decoder adopt a typical structure of Transformer, which includes three encoding and decoding layers with 8 heads. The channel encoder and decoder are composed of dense layers. More specifically, the network and training parameters are set as $B=128$, $L=30$, $V=128$, $N=16$, and $\eta=10^{-4}$ \cite{xie2021deep}. Besides, 1-gram and 3-grams are frequently utilized for calculating BLEU. In this letter, the weights of 1-gram are set as $u_1 = 1$ and $u_n = 0$ for $n \ne 1$, and the weights of 3-grams are assigned with $u_3 = 1$ and $u_n = 0$ for $n \ne 3$.


%


Figs. \ref{fig1} and \ref{3_gram} plot the BLEU scores of 1-gram and 3-grams versus the average SNR $\gamma_T \triangleq \mu {d_B}^2 P/\mathcal N$. For benchmarking comparison, the proposed DeepSSC without training Phase II is adopted as a baseline (labelled as ``No-II''). Furthermore, the integrated loss function in \eqref{eqn:inte} is adopted as another baseline (labelled as ``Int.''), where $w_1=w_2=1$.
It is clearly seen from both figures that the BLEU score of Bob is higher than that of Eve for ``No-II''. This is due to the fact that the average power gain of the main channel is larger than that of the eavesdropping channel. Moreover, it is observed from both figures that the proposed DeepSSC significantly reduces the BLEU score of Eve by comparing to ``No-II'', albeit at the cost of a slight degradation of reliability. {In some sense, a lower BLEU score at Eve represents a higher security. Furthermore, it is shown in Fig. \ref{fig1} that the BLEU score of Eve under DeepSSC diminishes with SNR. This demonstrates that the eavesdropping capability of Eve is substantially suppressed especially in the high SNR regime. This justifies the superiority of the proposed DeepSSC.} In addition, Figs. \ref{fig1} and \ref{3_gram} show that the proposed two-phase design surpasses the integrated one in terms of BLEU score. In other words, the proposed two-phase method can reach an improved security and reliability when compared to the integrated design. It should be noticed that this result may be not inapplicable to a general setting of $w_1$ and $w_2$.

Fig. \ref{fig2} presents the S-BLEU score versus the average SNR $\gamma_T$. It can be observed that the S-BLEU score of the baseline scheme generally decreases with the SNR. This is consistent with our intuition that more confidential semantic information is intercepted by Eve at high SNR. However, this decreasing tendency cannot be observed at low SNR, as eavesdropping becomes more challenging under such conditions. 
In addition, it is evident that the S-BLEU score of the proposed DeepSSC increases with SNR, indicating a noticeable improvement in the security of legitimate users provided by DeepSSC. Furthermore, Fig. \ref{fig2} demonstrates that the proposed method performs better than the integrated design in terms of S-BLEU.  {Nevertheless, since the S-BLEU score represents the difference between the BLEU scores at Bob and Eve, it can be seen from Fig. \ref{fig2} that the ``No-II'' scheme has the maximum S-BLEU score as SNR varies.}
\vspace{-10pt}




\section{Conclusions}\label{sec:con}
The DeepSSC has been proposed to provide secure semantic communications. To balance the tradeoff between security and reliability, the training process of neural networks has been divided into two phases. Particularly, Phases I and II have been proposed to guarantee reliability and security, respectively. The corresponding loss functions in two phases have been constructed according to Shannon capacity and secure channel capacity, which are approximated with variational inference. Moreover, the metric of BLEU has been used to evaluate the reliability of semantic communications and the metric of S-BLEU has been defined to assess the security of semantic communications. At last, simulation results have revealed that the DeepSSC yields a remarkable gain in semantic security especially at high SNR, albeit at the cost of a little bit inferior reliability.
\vspace{-10pt}

\bibliographystyle{IEEEtran}
 	\bibliography{ref1}

\end{document}